 \documentstyle[12pt]{article}
 \newlength{\dinwidth}
 \newlength{\dinmargin}
 \def\3{\ss}

\def\thebibliography#1{{\bf\noindent References}\markboth
 {REFERENCES}{REFERENCES}\list
 {[\arabic{enumi}]}{\settowidth\labelwidth{[#1]}\leftmargin\labelwidth
 \advance\leftmargin\labelsep
 \usecounter{enumi}}
 \def\newblock{\hskip .11em plus .33em minus -.07em}
 \sloppy
 \sfcode`\.=1000\relax}

\def\bbbone{{\mathchoice {\rm 1\mskip-4mu l} {\rm 1\mskip-4mu l}
{\rm 1\mskip-4.5mu l} {\rm 1\mskip-5mu l}}}
 \newcommand{\subs}[1]{\mbox{$\!$\scriptsize\it#1}}
 \newcommand{\Figure}[1]{\addtocounter{figure}{1}
                        {\small{\noindent{\bf Fig.\,\thefigure}\ \ #1}}}
 \newcommand{\Table}[2]{\addtocounter{table}{1}\newline%
                        {\small{\noindent{\bf Table\,\thetable}\ \ #1}}%
                         \begin{table}[tbh]\begin{minipage}{17cm}{}%
                         \begin{center}{\small #2}\end{center}%
                         \end{minipage}\end{table}}%
 \renewcommand{\thesection}{\arabic{section}.}
 \newcommand{\Section}[1]{\addtocounter{section}{1}%
                          \setcounter{subsection}{0}%
                          \setcounter{subsubsection}{0}%
                          {\bf\noindent \thesection\ #1}\newline\indent}
 \newcommand{\Sectionnoindent}[1]{\addtocounter{section}{1}%
                          \setcounter{subsection}{0}%
                          \setcounter{subsubsection}{0}%
                          {\bf\noindent \thesection\ #1}\newline}
 \renewcommand{\thesubsection}{\thesection\arabic{subsection}.}
 \newcommand{\Subsection}[1]{\addtocounter{subsection}{1}%
                             \setcounter{subsubsection}{0}%
                             {\small\bf\noindent \thesubsection\ #1}%
                             \newline\indent}
 \renewcommand{\thesubsubsection}{\thesubsection\arabic{subsubsection}.}
 \newcommand{\Subsubsection}[1]{\addtocounter{subsubsection}{1}%
                                {\noindent\thesubsubsection\it\ #1%
                                .}}
 \newcommand{\A}{{\cal A}}
 \newcommand{\Cstar}{C^{\ast}}
 
 \newcommand{\Dirac}{\mbox{$\not\!\!D$}}
 \newcommand{\Dm}{\triangle m^2}
 \newcommand{\Nc}{N_{\subs{c}}}
 \newcommand{\Tr}{\mbox{\rm Tr\,}}
 \newcommand{\bpl}{\mbox{{\large\bf (}}}
 \newcommand{\bpr}{\mbox{{\large\bf )}}}
 \newcommand{\eq}[1]{(\ref{#1})}
 \newcommand{\esk}{\enspace ,}
 \newcommand{\esp}{\enspace .}
 \newcommand{\lsim}{\raisebox{-3pt}{$\stackrel{<}{\sim}$}}
 \newcommand{\mcr}{m_{\subs{cr}}^2}
 \newcommand{\phin}{\phi^{(n)}}
 \newcommand{\phint}{\tilde{\phi}^{(n)}}
 \newcommand{\rn}{r^{(n)}}
 \newcommand{\vpn}{\varphi^{(n)}}
 
\begin{document}

 \noindent{\tt DESY $92-108$ \hfill ISSN $0418-9833$}\\
 {\tt July 1992}
 \renewcommand{\thefootnote}{{\protect\fnsymbol{footnote}}}

 \begin{center}
 \vspace{4cm}

 {\LARGE  Improving Multigrid and Conventional Relaxation} \\
 \medskip
 {\LARGE  Algorithms for Propagators
          \footnote{Work supported by Deutsche Forschungsgemeinschaft.}}

 \bigskip (Revised version)

 \addtocounter{footnote}{5}
 \vspace{3cm}
         Thomas Kalkreuter \footnote{E-mail: I02KAL@DSYIBM.DESY.DE} \\
         \smallskip
    {\em II. Institut f\"ur Theoretische Physik
         der Universit\"at Hamburg, \\
         Luruper Chaussee 149, W-2000 Hamburg 50, Germany}

 \vspace{1.5cm}\mbox{}

 \end{center}

 \vfill

 \begin{abstract}
  Practical modifications of deterministic multigrid and
  conventional relaxation algorithms are discussed.
  New parameters need not be tuned but are determined by the algorithms
  themselves.
  One modification can be thought of as ``updating on a last
  layer consisting of a single site''.
  It eliminates critical slowing down in computations
  of bosonic and fermionic propagators in a fixed volume.
  Here critical slowing down means divergence of asymptotic relaxation
  times as the propagators approach criticality.
  A remaining volume dependence is weak enough in case of bosons
  so that conjugate gradient can be outperformed.
  However, no answer can be given yet if the same is true for
  staggered fermions on lattices of realizable sizes.
  Numerical results are presented for propagators of bosons and of
  staggered fermions in 4-dimensional $SU(2)$ gauge fields.
 \end{abstract}

 \vfill\mbox{}

 \renewcommand{\thefootnote}{\arabic{footnote})}
 \setcounter{footnote}{0}
 \newpage

 \Section{Introduction}
 In Monte Carlo simulations of lattice gauge theories with fermions
 the most time-consuming part is the computation of the gauge field
 dependent fermion propagators.
 Conjugate gradient (CG) or minimal residual (MR) algorithms are state
 of the art \cite{OldAlg}.
 Great hopes to do better are attached to multigrid (MG) methods
 [2--10].
 In Ref.\ \cite{KalPL} the first MG computations without critical
 slowing down (CSD) in non-Abelian gauge fields (4-$d$ $SU(2)$)
 were presented.
 They prove that {\em the MG method can cope with the frustration
 which is inherent in non-Abelian gauge fields}.
 However, elimination of CSD succeeded only when an ``optimal''
 interpolation kernel \cite{KalPL,MacKalPalSpe} was used.
 The use of this optimal kernel for production runs is impractical
 because of computational complexity and storage space requirements.

 In this letter practical modifications of MG and
 conventional relaxation algorithms for propagators are discussed.
 A propagator $\phi$ is the solution of a linear equation
  \begin{equation}
    D \phi = f
  \label{propagator}
  \end{equation}
 on a $d$ dimensional lattice $\Lambda$ of sites $z$, for given $f$.
 In our case, $D = -\Delta + m^2$ for bosons, and $D = -\Dirac^2 + m^2$
 for fermions, where $\Delta$ and $\Dirac$ are the gauge covariant
 Laplace or Dirac operators (with periodic boundary conditions)
 respectively.
 Color indices are always suppressed, and $\phi (z)$ is an $\Nc
 \times \Nc$ matrix where $\Nc$ is the number of colors.

 CSD in computations in a fixed volume can be eliminated by ``updating
 on a last layer consisting of a single site''.
 Given an approximation $\phin$ to $\phi$, this updating amounts to
 rescaling $\phin$ by an $\Nc \times \Nc$ matrix
 $\Omega$ (in case of bosons or Wilson fermions):
  \begin{equation}
    \phin (z) \mapsto \phin (z)\,\Omega \ \ \ , \ \ \
     \Omega = \bpl \phin , D \phin \bpr^{-1} \bpl \phin , f\bpr \esk
  \label{rescaleBOSE}
  \end{equation}
 where
  \begin{equation}
    \bpl \varphi , \psi \bpr\ \equiv\ \frac{1}{| \Lambda |}
     \sum_{z \in \Lambda} \varphi (z)^{\dagger} \psi(z) \esp
  \end{equation}

 We will discuss Eq.~\eq{rescaleBOSE} and its generalization for
 staggered fermions, and related modifications in Sec.~2.
 Numerical results for propagators of bosons and of staggered fermions
 in 4-dimensional $SU(2)$ gauge fields will be presented in Sec.~3.

 \Sectionnoindent{Improving algorithms from a variational point of view}
 \Subsection{Updating on a $1^d$ MG layer}
 It is well known that the CSD of conventional iterative algorithms
 for solving Eq.~\eq{propagator} depends only on $m^2$ and {\em not\/}
 on the lattice size $|\Lambda |$.
 There is only an implicit dependence on $|\Lambda |$ (and $\beta$)
 through the value of $\mcr$\,, where $-\mcr$ denotes the lowest
 eigenvalue of $-\Delta$ or $-\Dirac^2$.
 The dimension $d$ enters in the scaling relation for relaxation times
 only through the constant of proportionality.
 Therefore one continues to have CSD on a lattice of only $2^d$ sites,
 and it seems necessary to go to a $1^d$ lattice in order to eliminate
 the appearance of CSD.

 When we update on a $1^d$ sublattice, we make the replacement
  \begin{equation}
    \phin (z) \mapsto \phin (z) + \A (z) ( \Omega - \bbbone ) \esp
  \label{lastpoint}
  \end{equation}
 Here $\A$ denotes a kernel which interpolates directly from a
 $1^d$ sublattice to $\Lambda$\@.
 $\Omega - \bbbone$ is the error of $\phin$ represented at the last
 site.
 In the MG context, $\Omega - \bbbone = \bpl \Cstar , D \A \bpr^{-1}
 \bpl \Cstar , f - D \phin \bpr$, where $\Cstar$ is the adjoint of the
 restriction operator which averages to a $1^d$ layer.
 {}From the variational point of view (VPV) an algorithm is set up in
 such a way that the functional $K [ \phi ] = \frac{1}{2} < \phi , D
 \phi > \mbox{$- < \phi , f >$}\,\equiv\,\Nc^{-1}\,\Tr\left[ \frac{1}{2}
 \bpl \phi , D \phi \bpr - \bpl  \phi , f \bpr \right]$ is lowered as
  far as possible in every  iteration.
 This leads to $\Cstar = \A$.
 {}From the VPV the optimal $\A$ in \eq{lastpoint} equals $\phin$.
 Thus, we obtain Eq.~\eq{rescaleBOSE}.
 Considered from the VPV alone without thinking of MG, one obtains
 \eq{rescaleBOSE} by rescaling $\phin$ with a matrix $\Omega$ that is
 determined such that $K[\phin \Omega ]$ is as low as possible.

 In case of staggered fermions we have to consider that there are $2^d$
 different pseudoflavors \cite{KalMacSpe}.
 In the limiting case of a pure gauge the fermionic problem amounts to
 computing $2^d$ decoupled bosonic propagators.
 Hence for staggered fermions we replace \eq{rescaleBOSE} by
  \begin{equation}
    \phin (z) \mapsto \phin (z)\,\Omega (H(z)) \esk
  \label{rescaleFERMI}
  \end{equation}
 where $H(z)$ denotes the pseudoflavor of $z$.
 Now the expression for $\Omega (H)$ is more complicated than that
 given in \eq{rescaleBOSE}.
 In practice, we determine the $\Omega (H)$'s by solving one linear
 $\Nc^2 2^d \times \Nc^2 2^d$ system, or -- making use of the
 independence of the even and odd sublattices -- by solving two
 systems of a quarter of that size.

 \Subsection{Related improvements from a VPV}
 Some related modifications of (MG) relaxation algorithms will be
 discussed now.
 The new parameters are not tunable, they are all determined
 by the algorithms themselves.

 \Subsubsection{Modified MG correction updating step}
 In conventional MG approaches one considers updates of the form
 $\phin \mapsto \phin + \vpn$ where $\vpn$ is obtained by interpolation
 of an approximate solution of a residual equation on a coarser lattice.
 We propose to generalize this to
  \begin{equation}
    \phin (z) \mapsto \phint \equiv
     \phin (z)\,\Omega + \vpn (z)\,\Theta \esp
  \label{improvedcorr}
  \end{equation}
 The two $\Nc\times\Nc$ matrices $\Omega$ and $\Theta$ are
 chosen such that $K [ \phint ]$ is minimized.
 In particular, this proposal may be an improvement in algorithms
 where the residual equation is only solved approximately,
 or in algorithms were coarse grid operators are not defined through
 the Galerkin prescription.
 For staggered fermions $\Omega$ and $\Theta$ in \eq{improvedcorr}
 become pseudoflavor dependent.

 \Subsubsection{Modified checkerboard SOR}
 Consider for illustration the bosonic problem.
 When we update at the even sites, we propose to modify SOR according to
  \begin{eqnarray}
    \phin (z)   \mapsto   \phin (z)\,\Omega & +\ \ \vpn (z)\,\Theta &
              \mbox{if $z$ is even} \esk \nonumber \\
    \phin (z)   \mapsto   \phin (z)\,\Xi    &                       &
              \mbox{if $z$ is odd} \esk
  \label{SOR}
  \end{eqnarray}
 where $\vpn (z) = ( 2d + m^2)^{-1} [ f(z) + \sum_{z'\mbox{\scriptsize
 n.n.}z} U ( z , z' ) \phin ( z' ) ]$, and  $U( z , z' )$ is the gauge
 field on the link $( z , z' )$.
 Again, the matrices $\Omega$, $\Theta$ and $\Xi$ should be chosen
 such that the functional $K$ gets minimized.
 The proposal \eq{SOR} expresses the view that in gauge theories one
 should have relaxation matrices rather than relaxation parameters that
 are real numbers.

 \Subsubsection{Modified damped Jacobi relaxation}
 Damped Jacobi relaxation can be generalized according to
 \eq{improvedcorr} with $\vpn = (2d + m^2)^{-1}\,\rn$.
 If one fixes $\Omega = \bbbone$, one recovers the MR algorithm
 that was used by Hulsebos et al.\ \cite{HSVLAT90,Vin,VinLAT91}.

 Finally we note that the proposals \eq{rescaleBOSE}, \eq{rescaleFERMI},
 \eq{improvedcorr} and \eq{SOR} respect gauge covariance.
 Iterative algorithms are gauge covariant in the sense that all $\phin$
 are gauge transformed by $g$ if $g$ is applied before relaxation is
 started.
 $\Omega$ is gauge invariant (or transforms like a matter field sitting
 at site $w$ in the adjoint representation, i.\ e.\ $\Omega \mapsto g_w
 \Omega g_w^{-1}$, when $f \rightarrow \delta_{z,w}$) etc.

 \clearpage
 \Section{Results for propagators in 4-dimensional
          {\boldmath $SU$}(2) gauge fields}
 It is well known \cite{OldAlg} that the convergence rate of iterative
 algorithms for propagators is governed by the condition number%
 \footnote{i.\ e.\ the ratio of the largest to the smallest eigenvalue}
 of $(-\Delta + m^2)$ or $(-\Dirac^2 + m^2)$.
 Therefore the CSD scaling relation for relaxation times $\tau$ reads
 \begin{equation}
   \tau \propto (\Dm)^{-z/2}  \enspace \mbox{for small} \enspace
    \Dm =  m^2 - \mcr \esk
  \label{scalingtau}
 \end{equation}
 where $z$ denotes the critical exponent.
 $\tau$ is defined by the {\em asymptotic\/} exponential decay of the
 norm of the residual.
 We measure all $\tau$'s and number of iterations in units
 of cycles which involve only one sweep through the finest lattice.
 The variational MG method used for solving Eq.~\eq{propagator} is
 described in some detail in Refs.\ \cite{KalMacSpe,KalPL,MacKalPalSpe}.
 Its implementation is actually a twogrid algorithm where the residual
 equation is solved exactly by CG.
 A scale factor of 3 is chosen in blocking.
 We use the gauge covariant ground-state projection MG method.
 An efficient algorithm for computing averaging kernels $C$ was
 described in Ref.\ \cite{KalNP}, and was used in this work.
 The (non-optimized) implementations of the MG programs require a
 factor of $3.2$/$7.8$ (bosons/staggered fermions) more arithmetic
 operations than CG when updating on a $1^d$ sublattice is included.
 Without that inclusion the factor is $2.1$/$4.5$.
 The amount of work for $C$ on the whole lattice is equivalent to less
 than 20 CG iterations.

 \Subsection{Propagators in pure gauges}
 In pure gauges the bosonic and fermionic problems are equivalent.
 There MG does a perfect job, CSD is completely eliminated with short
 $\tau$'s and $z=0$, Ref.\ \cite{KalPL}.
 Rescaling \eq{rescaleBOSE} does not improve the performance any
 further.
 The improved correction scheme \eq{improvedcorr}, however, is able
 to halve $\tau$ and to bring it close to 1.
 Combining \eq{rescaleBOSE} with conventional 1-grid relaxation
 causes CSD (in the sense stated above) to disappear.
 This is shown in Fig.~1.
 The norm of the residual is not monotonically decreased for small
 $m^2$.
 This feature also shows up in CG which minimizes $K$
 (i.\ e.\ the residual $r$ in the norm induced by the scalar product
 $<\,\cdot\,,D^{-1}\,\cdot\,>$) rather than $|| r ||$ itself.
 This is an interesting point:
 Instead of~\eq{rescaleBOSE}, one could think of rescaling $\phin$ by
 another matrix $\Omega'$ which is chosen such that $|| \rn ||$ is
 minimized.
 However, in this case there is no difference to 1-grid relaxation.
 This remains true in nontrivial gauge fields.
 Finally we note that practically $\Omega = \bbbone$ as soon as
 $|| r ||$ decays exponentially.
 Then the step \eq{rescaleBOSE} could be switched off.
 This statement holds also in nontrivial gauge fields and for MG.

 \Subsection{Bosonic propagators in nontrivial gauge fields}
 The remarks made above about 1-grid relaxation plus
 \eq{rescaleBOSE} apply in nontrivial gauge fields as well.
 Fig.~1 looks the same when $m^2$ is replaced by $\Dm$.
 MG plus \eq{rescaleBOSE} beats CSD, too.
 Compared with CG and 1-grid plus \eq{rescaleBOSE}, this
 modified MG performs better the more critical the system is.
 Convergence for $\Dm = 10^{-6}$ on $12^4$ and $18^4$ lattices is shown
 in Fig.~2.%
 \footnote{Using SOR as a smoother contradicts the conventional MG
           wisdom.
           However, in nontrivial gauge fields any over-relaxation
           yields better performance than Gauss-Seidel and much better
           performance than damped Jacobi or MR.
           A similar result was reported in a recent paper on MG
           gauge fixing \cite{HulLauSmi}.}
 It must be emphasized that $\Omega$ really needs to be a matrix.
 Simple rescaling of $\phin$ with a real number does not succeed
 in eliminating CSD.%
 \footnote{A remark about 1-grid plus \eq{rescaleBOSE} is in order here:
           It can be proved by induction that in $SU(2)$ gauge fields
           $\Omega \propto \bbbone$ always holds if one starts
           relaxation with $\phi^{(0)} = 0$.
           However, $\Omega \propto \bbbone$ is not true in general.}
 The improved MG correction scheme \eq{improvedcorr} also beats CSD.
 However, at finite $\beta$ it is not able to halve $\tau$'s.
 Using the modified SOR version \eq{SOR} does not pay.
 If one fixes $\Xi = \bbbone$, one has nothing else but conventional
 Gauss-Seidel relaxation.
 Retaining $\Xi$ yields little difference in performance.

 \Subsection{Propagators of staggered fermions in nontrivial gauge
             fields}
 In case of staggered fermions $\mcr$ is much closer to zero than
 in case of bosons.
 Actually, it is often assumed that the finite value of $\mcr$ can be
 neglected completely in \eq{scalingtau} so that $\tau \propto m^{-z}$.
 However, this neglect of $\mcr$ is not justified on lattices amenable
 in size to date.
 For bosons, scaling \eq{scalingtau} without violations
 is observed for $\Dm \lsim 0.01$, Ref.\ \cite{KalPL}.
 For staggered fermions it was verified that the scaling relation
 \eq{scalingtau} also holds.
 The critical exponent $z$ equals 2 for conventional relaxation (fixed
 $\omega$) and for MG without \eq{rescaleFERMI}.
 It was also verified that the constant of proportionality in
 \eq{scalingtau} is independent of the lattice size.
 This statement is true for 1-grid as well as for MG algorithms.
 To obtain these results requires a careful analysis of data.
 The subtle point is that for fermions \eq{scalingtau} is obeyed only
 for $\Dm \lsim 0.001$, and asymptotic decay in the sense
 that only the slowest mode governs convergence does not set in before
 400 -- 500 iterations in 4-dimensional $SU(2)$ gauge fields.

 The lesson is that $\mcr$ must not be neglected, at least for lattices
 up to $18^4$.
 In practice this might mean that on such relatively small lattices
 $m^2$ must also be given small negative values in order to
 study effects of CSD.
 This is of course artificial, but it is necessary when one wants to
 obtain results for $z$ which are reliable for predictions how
 algorithms perform on large lattices.
 If one does not investigate systems close enough to criticality
 and if one does not take care that decay rates are really asymptotic,
 there is the danger of extracting wrong values for $z$, even in case
 of pure gauges.

 If one is interested in the question how well algorithms perform on
 lattices of sizes that are used in present day simulations,
 the foregoing discussion might appear too academic.
 It might appear more natural to ask how many iterations
 are needed to obtain a given accuracy.
 {}From this viewpoint it turned out that MG is competitive or even
 superior to conventional algorithms in 2-$d$ models
 \cite{BenBraSol,BERV}.
 However, it is more difficult to reach decisive conclusions in $d=4$.
 Preliminary results on $16^4$ \cite{VinLAT91} and $18^4$
 \cite{KalLeipzig} lattices (both at $\beta = 2.7$)
 indicated that the MG methods tested so far will not be able to
 outperform CG.
 But we expect that the situation will be different on larger lattices.
 Details will be reported elsewhere \cite{Kalthesis}.

 We carry on with results of the modifications proposed in this article.
 Table~1 gives a survey of convergence on $12^4$ and $18^4$ lattices.
 Rescaling \eq{rescaleFERMI} does not pay for positive $m^2$ on small
 lattices.
 But including \eq{rescaleFERMI} in algorithms brings $z$ down to zero,
 i.\ e.\ CSD is eliminated.
 An analogue of Fig.~2 for staggered fermions is given with Fig.~3.
 Mind, however, that relaxation algorithms for fermions are used with
 lexicographic, not checkerboard, updating.

 We conclude this section with remarks on MR.
 Since MR can be viewed as an optimized Jacobi relaxation, it comes as
 no surprise that its convergence properties are worse than those of
 SOR.
 Working with pseudoflavor dependent matrices leads to no practical
 improvement.
 Over-relaxed MR versions have not been investigated yet.

 \Subsection{Cautionary Remark}
 To the author's knowledge there exists no study in the literature
 where $\mcr$ is not disregarded in case of staggered fermions.
 This neglect is only justified by the smallness of $\mcr$ but it has
 never been checked whether the neglect is justified.
 A result of the present work is the validity of the relation $\tau =
 \mbox{const}/\Dm$ in 1-grid and variational MG relaxation, with a
 constant which is {\em independent of the lattice size\/}.
 Therefore the study of the {\em asymptotic} behavior of $\tau$
 when the linear extension of the lattice and $1/\triangle m$ are
 changed proportionally, can be determined from studies at fixed volume.
 Elimination of the $1 / \Dm$ divergence on a lattice of fixed size
 implies the absence of CSD in computations where all quantities are
 scaled appropriately.
 Certainly, in physical applications (Monte Carlo simulations) the
 inverse mass should be smaller than the extension of the lattice,
 but this is an aspect of finite size effects on physical observables.

 By neglecting $\mcr$ and trying to determine $z$ under scaling
 conditions, one can at most obtain some effective $z_{\subs{eff}}$.
 This $z_{\subs{eff}}$ contains however a great deal of arbitrariness
 and cannot be defined uniquely.
 One can run into difficulties with this procedure \cite{HSVLAT90}.
 The author admits that a $z_{\subs{eff}}$ is of more practical
 relevance as long as numerical simulations are limited to
 lattice sizes where $m$ is not really small.
 But we look for algorithms which can be used in future large scale
 computations, and for these it will be $z$ and not $z_{\subs{eff}}$
 which governs CSD\@.
 There is one weak point in this reasoning, and that is the remaining
 volume effect of \eq{rescaleFERMI}.
 $z=0$ is valid asymptotically, but it takes longer to reach the
 asymptotic regime the larger the lattice becomes.
 Therefore one might have to go back to a $z_{\subs{eff}}$, but this is
 an open question.

 \Section{Conclusions}
 Updating on a last $1^d$ MG layer provides an astonishingly simple
 modification which eliminates CSD of {\em asymptotic} relaxation
 times in MG and even in 1-grid relaxation algorithms for propagators.
 As soon as the decay of the error is exponential, updating on the
 $1^d$ sublattice can be switched off.
 Therefore additional work must only be invested in the initial and
 in an intermediate stage of computations.
 Since the modified algorithms have $z = 0$, we expect that CG will
 eventually be outperformed.
 What remains, however, is a volume dependence on how fast the
 asymptotic regime is reached.
 For bosons it was shown that CG is outperformed.
 But we feel unable to predict whether the same methods will pay for
 staggered fermions on lattices of realizable sizes.

 \smallskip\noindent{\small {\sc Acknowledgments}
 \newline\indent
 The investigation of ``updating on a last $1^d$ MG layer'' was
 motivated by the observation of Gerhard Mack that this eliminates CSD
 in a simple toy model.
 I am indebted to him for stimulating discussions.
 I would also like to thank S.\,Meyer for his interest in this work
 and for discussions.
 Financial support by Deutsche Forschungsgemeinschaft
 is gratefully acknowledged.
 For providing resources, advice and help I wish to thank
 HLRZ J\"ulich and its staff.
 }


 {\noindent\bf Table}
 \Table{Convergence in  computations of propagators of staggered
        fermions with $f(z) = \delta_{z,0}$ in 4-$d$ $SU(2)$
        gauge fields.
        Given is the number of iterations necessary for reducing
        $\ln || r^{(0)} ||$ by 10.
        All propagators are initialized with zero.
        1-grid and MG SOR are swept in lexicographic ordering.}
 {\begin{tabular}{|l||r|r|r|r|r|r|}
 \hline
 \multicolumn{7}{|c|}{pure gauge configurations ($\beta = \infty$)} \\
 \hline
               & \multicolumn{6}{|c|}{$m^2 =$} \\
 algorithm and lattice size
               & $10^{-1}$ & $10^{-2}$ & $10^{-3}$ & $10^{-4}$ &
                 $10^{-5}$ & $10^{-6}$ \\
 \hline\hline
 CG on $12^4$  & $17$  & $17$  & $17$  & $17$  & $17$  & $17$ \\
 \hline
 CG on $18^4$  & $28$  & $30$  & $33$  & $35$  & $37$  & $39$ \\
 \hline\hline
 1-grid SOR plus \eq{rescaleFERMI}, $\omega = 1.90$, on $12^4$
               & $105$ & $125$ & $155$ & $195$ & $230$ & $270$\\
 \hline
 1-grid SOR plus \eq{rescaleFERMI}, $\omega = 1.90$, on $18^4$
               & $110$ & $125$ & $155$ & $195$ & $230$ & $270$\\
 \hline\hline
 MG SOR without \eq{rescaleFERMI}, $\omega = 1.09$, on $12^4$
               & $21$ & $23$ & $23$ & $23$ & $23$ & $23$\\
 \hline
 MG SOR without \eq{rescaleFERMI}, $\omega = 1.09$, on $18^4$
               & $20$ & $23$ & $23$ & $23$ & $23$ & $23$\\
 \hline\hline
 \multicolumn{7}{|c|}{nontrivial gauge fields ($\beta = 2.7$)} \\
 \hline
               & \multicolumn{6}{|c|}{$\Dm =$} \\
 algorithm and lattice size
               & $10^{-1}$ & $10^{-2}$ & $10^{-3}$ & $10^{-4}$ &
                 $10^{-5}$ & $10^{-6}$ \\
 \hline\hline
 CG on $12^4$  & $65$  & $175$ & $265$ & $300$ & $320$ & $340$\\
 \hline
 CG on $18^4$  & $65$  & $180$ & $350$ & $415$ & $455$ & $495$\\
 \hline\hline
 1-grid SOR plus \eq{rescaleFERMI}, $\omega = 1.90$, on $12^4$
               & $100$ & $130$ & $530$ & $560$ & $630$ & $710$\\
 \hline
 1-grid SOR plus \eq{rescaleFERMI}, $\omega = 1.90$, on $18^4$
               &  $95$ & $125$ & $710$ & $1090$& $1320$& $1540$\\
 \hline\hline
 MG SOR plus \eq{rescaleFERMI}, $\omega = 1.96$, on $12^4$
               & $180$ & $185$ & $385$ & $425$ & $475$ & $535$\\
 \hline
 MG SOR plus \eq{rescaleFERMI}, $\omega = 1.96$, on $18^4$
               & $180$ & $185$ & $530$ & $775$ & $950$ & $1070$\\
 \hline\hline
 \end{tabular}}

 {\noindent\bf Figure captions}

 \Figure{1-grid SOR plus \eq{rescaleBOSE} eliminates CSD, shown here
         for a pure gauge.
         ($\omega\!=\!1.90$, checkerboard updating)
         The 6 curves correspond to $m^2 = 0.1, 0.01, \ldots ,
         10^{-6}$ on a $12^4$ (non-staggered) lattice with $m$
         decreasing from left to right.
         A small volume effect remains, not for $\tau$ (i.\ e.\ the
         asymptotic decay rate) but with respect to how fast the
         asymptotic regime is reached: e.\ g.\ on an $18^4$ lattice the
         number of iterations needed to obtain a given accuracy of $||
         r ||$ for $m^2 = 10^{-6}$ is increased by $\sim 20$.
         The figure looks the same for bosons in nontrivial gauge
         fields when $m^2$ is replaced by $\Dm$.}

 \Figure{Convergence for bosonic propagators with $\Dm = 10^{-6}$ in
         quenched 4-$d$ $SU(2)$ gauge fields equilibrated
         with Wilson's action at $\beta = 2.7$.
         The numbers refer to the following algorithms:
         1/2: variational MG SOR ($\omega = 1.50$) plus
         \eq{rescaleBOSE} on a $12^4$/$18^4$ lattice;
         3/4: 1-grid SOR ($\omega = 1.91$) plus
         \eq{rescaleBOSE} on a $12^4$/$18^4$ lattice;
         5/6: CG on a $12^4$/$18^4$ lattice.
         Relaxation algorithms are swept in checkerboard fashion.
         The critical masses are $\mcr = -0.7726281$/$-0.7554339$.
         Without \eq{rescaleBOSE} MG SOR and 1-grid SOR have $\tau$'s
         of $O(10^5)$ \cite{KalPL}.}

 \Figure{Convergence for propagators of staggered fermions with $\Dm =
         10^{-6}$ in quenched 4-$d$ $SU(2)$ gauge fields at $\beta =
         2.7$.
         The numbers refer to the following algorithms:
         1/2: variational MG SOR ($\omega = 1.96$) plus
         \eq{rescaleFERMI} on a $12^4$/$18^4$ lattice;
         3/4: 1-grid SOR ($\omega = 1.90$) plus
         \eq{rescaleFERMI} on a $12^4$/$18^4$ lattice;
         5/6: CG on a $12^4$/$18^4$ lattice.
         Relaxation algorithms are swept in lexicographic ordering.
         The critical masses are $\mcr = -0.0368447$/$-0.0096640$.
         Without \eq{rescaleFERMI} MG SOR and 1-grid SOR have $\tau$'s
         of $O(10^5)$.}


\bigskip

 \begin{thebibliography}{99}

 {\small
 \bibitem{OldAlg}
              C.B.\,Chalmers, R.D.\,Kenway and D.\,Roweth,
              J. Comp. Phys. 70 (1987) 500;\

              P.B.\,Mackenzie,
              Nucl. Phys. B (Proc. Suppl.) 17 (1990) 103;\

              D.\,Henty, R.\,Setoodeh and C.T.H.\,Davies,
              Nucl. Phys. B337 (1990) 487

 \bibitem{BMMR}
              R.C.\,Brower, K.J.M.\,Moriarty, E.\,Myers and C.\,Rebbi,
              in: Multigrid Methods, ed.\ S.F.\,McCor\-mick
              (Marcel Dekker, New York, 1988)


 \bibitem{BenBraSol}
              R.\,Ben-Av, A.\,Brandt and S.\,Solomon,
              Nucl. Phys. B329 (1990) 193;\

              R.\,Ben-Av, A.\,Brandt, M.\,Harmatz, E.\,Katznelson,
              P.G.\,Lauwers, S.\,So\-lo\-mon and  K.\,Wo\-lo\-wes\-ky,
              Phys. Lett. B253 (1991) 185;\
             %
              Nucl. Phys. B (Proc. Suppl.) 20 (1991) 102;\

              R.\,Ben-Av, P.G.\,Lauwers and S.\,So\-lo\-mon,
             Nucl. Phys. B 374 (1992) 249;\

              P.G.\,Lauwers and S.\,So\-lo\-mon,
              Int. J. Mod. Phys. C3 (1992) 149

 \bibitem{BRV}
              R.C.\,Brower, C.\,Rebbi and E.\,Vicari,
              Phys. Rev. D43 (1991) 1965;\
             %
              Phys. Rev. Lett. 66 (1991) 1263;\

              R.C.\,Brower, K.J.M.\,Moriarty, C.\,Rebbi and E.\,Vicari,
              Nucl. Phys. B (Proc. Suppl.) 20 (1991) 89;\
             %
              Phys. Rev. D43 (1991) 1974

 \bibitem{HSVLAT90}
              A.\,Hulsebos, J.\,Smit and J.C.\,Vink,
              Nucl. Phys. B (Proc. Suppl.) 20 (1991) 94;\
             %
              Int. J. Mod. Phys. C3 (1992) 161;\
             %
              Nucl. Phys. B368 (1992) 379

 \bibitem{BERV}
              R.C.\,Brower, R.G.\,Edwards, C.\,Rebbi and E.\,Vicari,
              Nucl. Phys. B366 (1991) 689

 \bibitem{KalMacSpe}
              T.\,Kalkreuter, G.\,Mack and M.\,Speh,
              Int. J. Mod. Phys. C3 (1992) 121

 \bibitem{Vin}
              J.C.\,Vink,
              Phys. Lett. B272 (1991) 81

 \bibitem{VinLAT91}
              J.C.\,Vink,
              Nucl. Phys. B (Proc. Suppl.) 26 (1992) 607

 \bibitem{KalPL}
              T.\,Kalkreuter,
              Phys. Lett. B276 (1992) 485

 \bibitem{MacKalPalSpe}
              G.\,Mack, T.\,Kalkreuter, G.\,Palma and M.\,Speh,
              {\em Effective Field Theories},
              preprint DESY 92--070,
              to appear in the proceedings of the 31st IUKT,
              Schladming, February 1992

 \bibitem{KalNP}
              T.\,Kalkreuter,
              Nucl. Phys. B376 (1992) 637

 \bibitem{HulLauSmi}
              A.\,Hulsebos, M.L.\,Laursen and J.\,Smit,
              {\em $SU(N)$ Multigrid Landau Gauge Fixing},
              KFA J\"ulich preprint HLRZ--92--15, May 1992

 \bibitem{KalLeipzig}
              T.\,Kalkreuter,
              {\em Multigrid for staggered fermions in 4-dimensional
                   $SU(2)$ gauge fields},
              talk presented at the DFG-Colloquium ``Dynamical
              Fermions'' held in Leipzig, March 1992

 \bibitem{Kalthesis}
              T.\,Kalkreuter,
              {\em Ph.D.\ thesis}, in preparation

 }
 \end{thebibliography}
 \end{document}